\documentclass[a4paper,11pt]{article}
\usepackage{pos}

\usepackage{subcaption}
\usepackage{units}
\usepackage{gensymb}
\usepackage{mathrsfs}
\usepackage{mathtools}
\usepackage{bm}

\def\Journal#1#2#3#4{{#1} {\bf #2}, #3 (#4)}

\def\AA{\em A\&A}
\def\ADASS{\em Proceedings of XXXI Astronomical Data Analysis Software and Systems (ADASS) conference}
\def\APJ{\em Astrophys. J.}
\def\GAL{\em Galaxies}
\def\ICRCA{\em Proceedings of $36^{\text{th}} $ International Cosmic Ray Conference (ICRC)}
\def\ICRCB{\em Proceedings of $37^{\text{th}} $ International Cosmic Ray Conference (ICRC)}
\def\ICRCC{\em Proceedings of $33^{\text{th}} $ International Cosmic Ray Conference (ICRC)}
\def\JCAP{\em JCAP}
\def\MORIOND{\em Proceedings of 2022 VHEPU session of the  $56^{\text{th}} $ Rencontres de Moriond}

\def\PDU{\em Physics of the Dark Universe}
\def\PLB{{\em Phys. Lett.}  B}
\def\PRL{\em Phys. Rev. Lett.}
\def\PRD{{\em Phys. Rev.} D}
\def\ZENODO{\em Zenodo}

\title{Combined search in dwarf spheroidal galaxies for branon dark matter annihilation signatures with the MAGIC Telescopes}
\ShortTitle{Multi-target branon dark matter search in dSphs with MAGIC}

\author*[a]{T. Miener}
\author[a]{D. Nieto}
\author[b,c]{V. Gammaldi}
\author[d]{D. Kerszberg}
\author[d]{J. Rico}
\onbehalf{on behalf of the MAGIC Collaboration}

\affiliation[a]{Instituto de F\'{i}sica de Part\'{i}culas y del Cosmos and Department of EMFTEL, Universidad Complutense de Madrid, Pza. Ciencias 1, E-28040 Madrid, Spain}
\affiliation[b]{Departamento de F\'{i}sica Te\'{o}rica, Facultad de Ciencias, Mod. 15,  Universidad Autónoma de Madrid, E-28049 Madrid, Spain}
\affiliation[c]{Instituto de F\'{i}sica Te\'{o}rica, UAM-CSIC, Calle Nicol\'{a}s Cabrera 13-15, Campus de Cantoblanco, E-28049 Madrid, Spain}
\affiliation[d]{Institut de F\'{i}sica d'Altes Energies (IFAE), The Barcelona Institute of Science and Technology (BIST) Campus UAB, Edifici Cn, E-08193 Bellaterra (Barcelona), Spain}

\emailAdd{tmiener@ucm.es}

\abstract{One of the most pressing questions for modern physics is the nature of dark matter (DM). Several efforts have been made to model this elusive kind of matter. The largest fraction of DM cannot be made of any of the known particles of the Standard Model (SM). We focus on brane world theory as a prospective framework for DM candidates beyond the SM of particle physics. The new degrees of freedom that appear in flexible brane world models, corresponding to brane fluctuations, are called branons. They behave as weakly interacting massive particles (WIMPs), which are one of the most favored candidates for DM. We present a multi-target DM search in dwarf spheroidal galaxies for branon DM annihilation signatures with the ground-based gamma-ray telescope MAGIC leading to the most constraining branon DM limits in the TeV mass range.}

\FullConference{%
  7th Heidelberg International Symposium on High-Energy Gamma-Ray Astronomy (Gamma2022)\\
  4-8 July 2022\\
  Barcelona, Spain\\}

\begin{document}
\maketitle

\section{Introduction}
The nature of dark matter (DM) is still an open question for modern physics. 
According to the \textit{Planck 2018 results}~\cite{Aghanim:2018eyx}, non-baryonic cold DM accounts for $84\%$ of the matter density of the Universe based on astrophysical and cosmological evidences. Brane-world theory as a prospective framework for DM candidates~\cite{2003PhRvL..90x1301C} proposes massive brane fluctuations (branons) as a natural TeV DM candidate, since their characteristics match the ones of weakly interacting massive particles (WIMPs)~\cite{2012PhRvD..86b3506S}.

Dwarf spheroidal galaxies (dSphs) are preferred targets for indirect DM searches because they are not expected to host strong conventional gamma-ray emitters that may hinder the detection of a subdominant DM signal. In addition, they are close by, and have high mass-to-light ratios. Finally, compared to other very prominent targets for DM searches the Galactic Center (GC) and galaxy clusters~\cite{2002PhRvL..88s1301M,2007ApJ...654..897B}, dSphs are spatially less extended. In this work, we are searching for branon dark matter annihilation signatures in dSphs with the MAGIC telescopes.

\section{Branon dark matter}

\begin{figure}[ht]
    \centering
    \begin{subfigure}{.49\textwidth}
        \includegraphics[width=\textwidth]{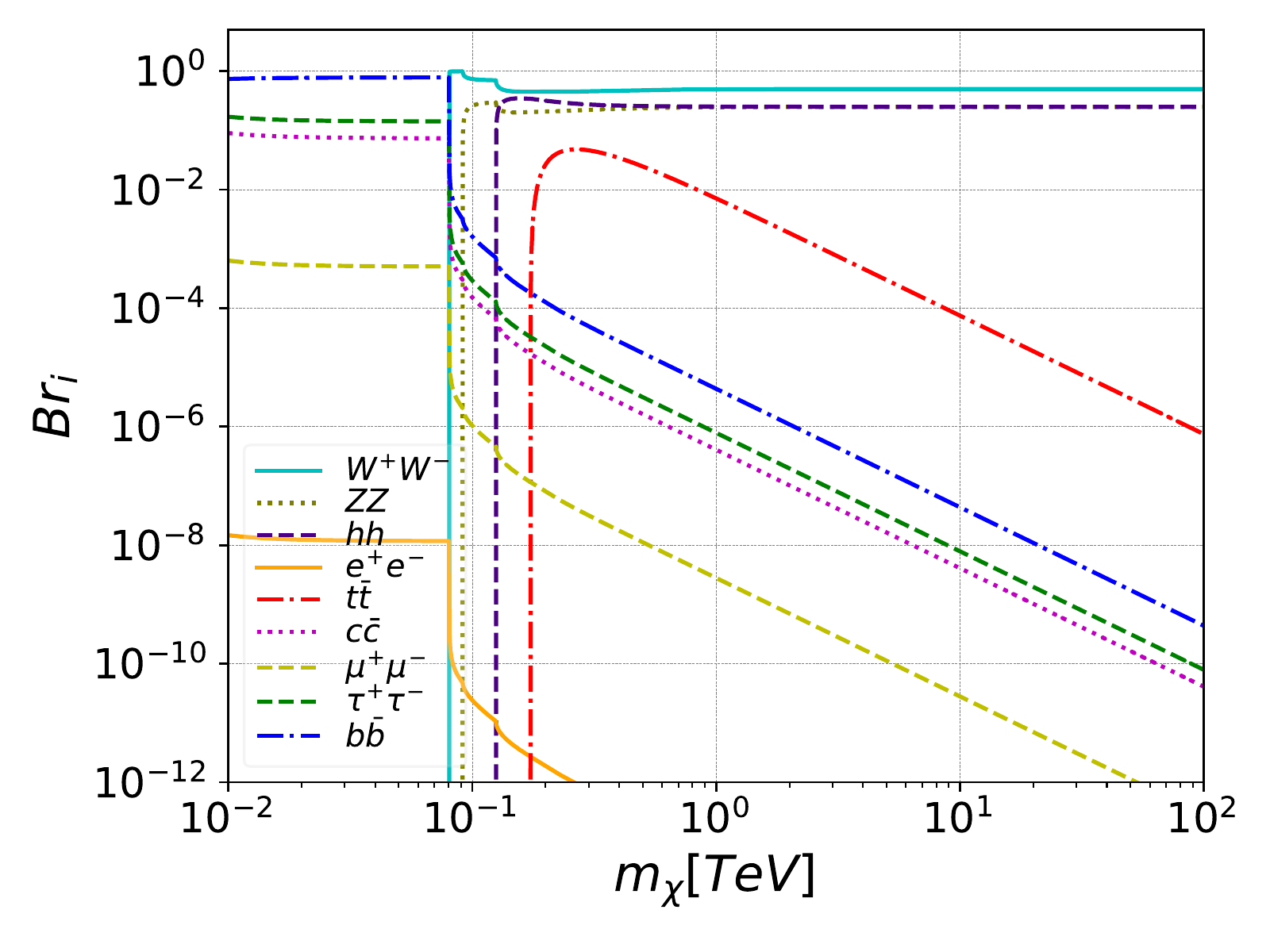}
    \end{subfigure}
    \begin{subfigure}{.49\textwidth}
        \includegraphics[width=\textwidth]{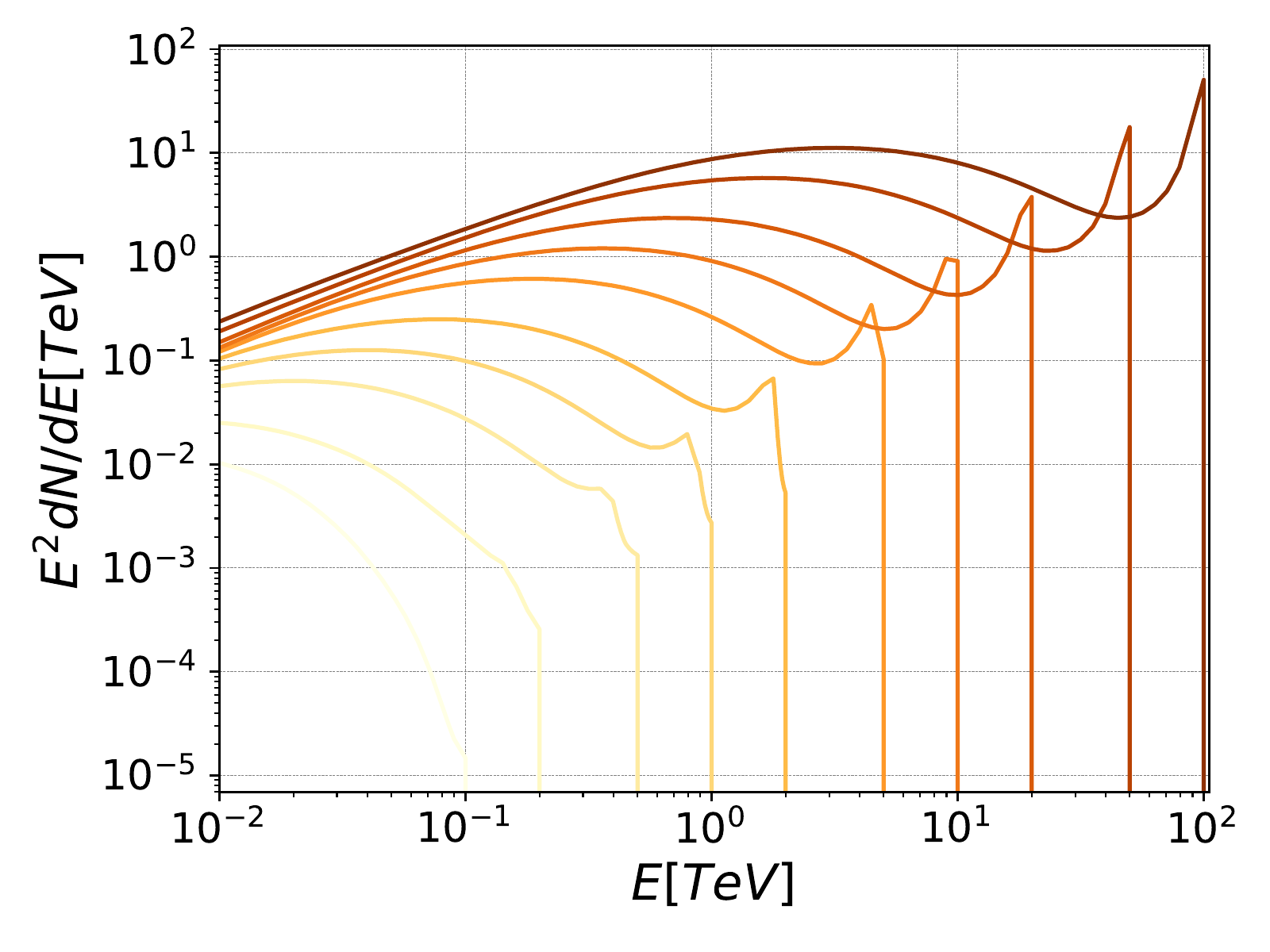}
    \end{subfigure}
\caption{Taken from~\cite{2022JCAP...05..05}. {\em Left:} The branon branching ratios as a function of $ m_{\chi} $ for DM masses from $ \unit[10]{GeV} $ up to $ \unit[100]{TeV} $. {\em Right:} The differential photon yield per branon annihilation $ \text{d}N_{\text{BDM}}/\text{d}E $ (Eq. \ref{eq:DifPhotonYield}) for a set of DM masses (from light to dark: 0.1, 0.2,
0.5, 1, 2, 5, 10, 20, 50 and $ \unit[100]{TeV} $).}
\label{fig:Branching_ratios_and_spectra}
\end{figure}

The expected photon flux produced by branon DM annihilation is composed of the \textit{astrophysical} factor (\textit{J}-factor), which depends on both the distance $ l $ and the DM distribution at the source region $ \rho_{\text{DM}} $, and the \textit{particle physics} factor, mainly the differential photon yield per branon annihilation. It can be expressed from a given solid angle region in the sky, $\Delta\Omega$, as
\begin{equation}
    \label{eq:Branon_Flux}
    \frac{\text{d}\Phi_{\text{BDM}}}{\text{d}E}\left( \langle\sigma v\rangle \right) = \underbrace{\int_{\Delta\Omega} d\Omega' \int_{\text{l.o.s.}} dl \, \rho_{\text{DM}}^{2} (l,\Omega')}_{\text{Astrophysics}} \cdot \underbrace{\frac{1}{4\pi} \frac{\langle\sigma v\rangle}{2m^{2}_{\chi}} \frac{\text{d}N_{\text{BDM}}}{\text{d}E}}_{\text{Particle physics}}
\end{equation}
\noindent
with
\begin{equation}
    \label{eq:DifPhotonYield}
     \frac{\text{d}N_{\text{BDM}}}{\text{d}E} = \sum_{i=1}^{n} \text{Br}_{i} \frac{\text{d}N_{i}}{\text{d}E},
\end{equation}
\noindent
where $ \langle\sigma v\rangle $ is the thermally-averaged annihilation cross section (our parameter of interest and therefore the only free parameter in our likelihood analysis of Sec.~\ref{sec:LklAnalysis}), $ m_{\chi} $ is the mass of the branon DM particle and $ \text{l.o.s.} $ stands for line-of-sight. The  differential photon yields per annihilation into SM pairs $ \text{d}N_{i}/\text{d}E $ are taken from the PPPC 4 DM ID distribution~\cite{2011JCAP...03..051C}. The left panel of Fig.~\ref{fig:Branching_ratios_and_spectra} shows the branon branching ratios $ \text{Br}_{i} $ as a function of $ m_{\chi} $~\cite{2022JCAP...05..05}. The differential photon yield per branon annihilation $ \text{d}N_{\text{BDM}}/\text{d}E $ is depicted for a set of DM masses in the right panel of Fig.~\ref{fig:Branching_ratios_and_spectra}.

\section{Dwarf spheroidal galaxies observations with the MAGIC telescopes}
\label{sec:Observation}

The \textit{Florian Goebel} Major Atmospheric Gamma-ray Imaging Cherenkov (MAGIC) telescopes\footnote{\href{https://magic.mpp.mpg.de/}{https://magic.mpp.mpg.de/}} are located at the Roque de los Muchachos Observatory ($ 28.8^{\circ}$ N, $ 17.9^{\circ}$ W) on the Canary Island of La Palma, Spain. MAGIC consists of two 17-m diameter reflector imaging atmospheric Cherenkov telescopes (IACTs), which inspect the very-high energy (VHE, $ \gtrsim \unit[50]{GeV} $) gamma-ray sky probing the most extreme astrophysical environments in our universe.

\begin{figure}[h]
    \centering
    \includegraphics[width=0.8\textwidth]{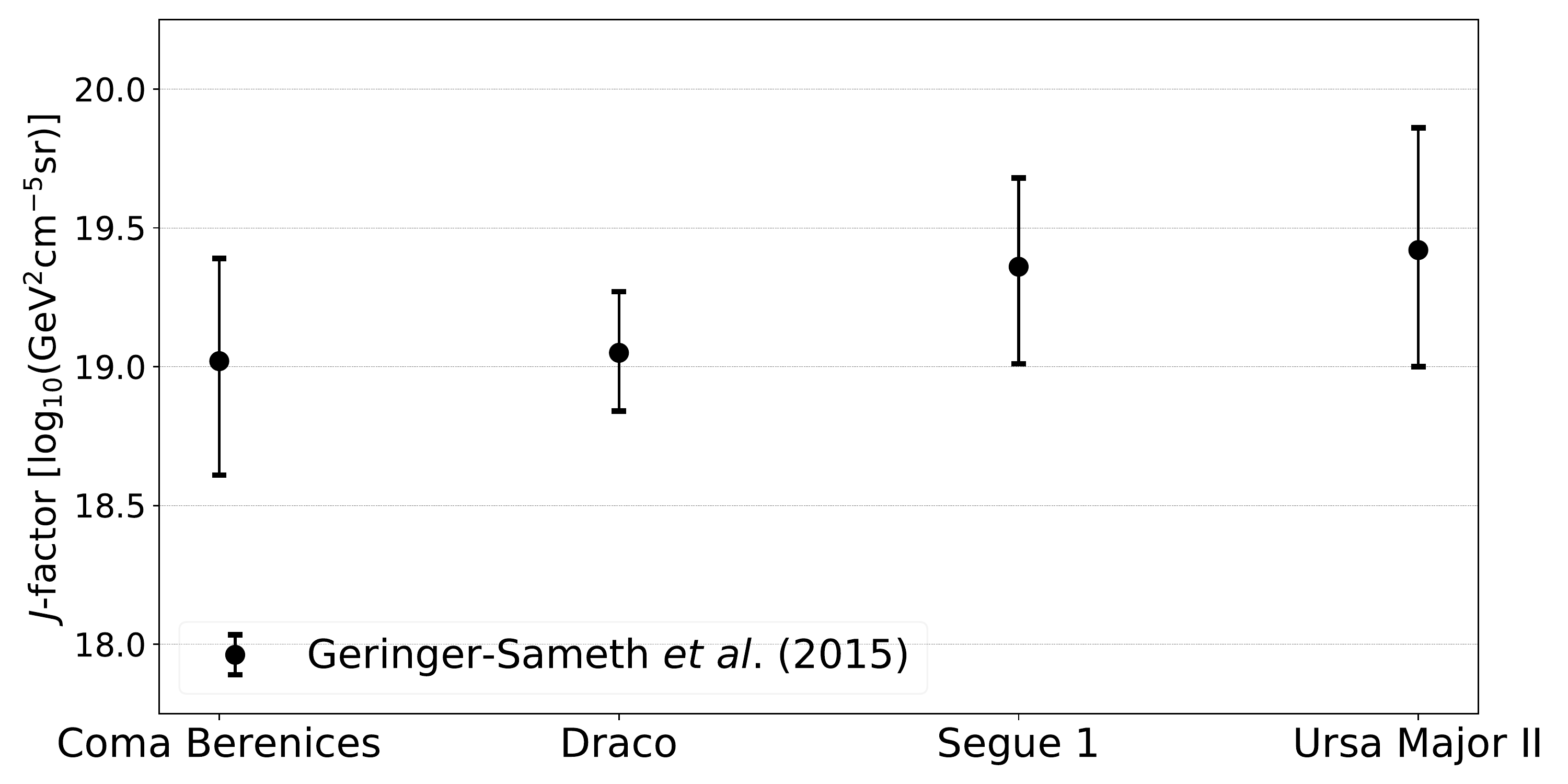}
\caption{The total \textit{J}-factor values and its uncertainties from GS15~\cite{2015ApJ...801...74G} for all considered dSphs.}
\label{fig:Jfactors}
\end{figure}

The MAGIC Collaboration has carried out extensive observations on dSphs in the Northern Hemisphere throughout the years, motivated by the search for DM signals in these objects. We include the dSph observations of Segue 1 (158h), Ursa Major II (95h), Draco (52h), and Coma Berenices (49h) with a total exposure of 354h in our work to align with the combined model independent DM search by MAGIC in~\cite{2022PDU....3500912A}. We are also using the total \textit{J}-factor and its statistical uncertainty from Geringer-Sameth \emph{et al.}~\cite{2015ApJ...801...74G} (GS15) as~\cite{2022PDU....3500912A}. Their corresponding values and its $\pm 1\sigma$ uncertainties are listed in Tab.~\ref{Tab:dSph} and visualized in Fig.~\ref{fig:Jfactors}.

\begin{table}[h]
\centering
\scriptsize
\begin{tabular}{ccccccccc}
\hline
\hline
Name & Distance & $l, b$ & $\log_{10}J$ & Zd & $ T_{\text{obs}} $ & E & $\theta$ & S$_{\text{Li\&Ma}}$ \\
& [kpc] &  [\degree] & [$\log_{10}(\rm{GeV}^2 \rm{cm}^{-5}\rm{sr})$] &  [\degree] & [h] & [TeV] & [\degree] & [$\sigma$]\\
\hline
Coma~Berenices & $44$ & $241.89,\: 83.61$ & $19.02^{+0.37}_{-0.41}$ & $5 - 37 $ & $49$ & $0.06 - 100$ & $0.17$ & $-$ \\
Draco & $76$ & $86.37,\: 34.72$ & $19.05^{+0.22}_{-0.21}$ & $29 - 45$ & $52$ & $0.07 - 100$ & $0.22$ & $-$ \\
Segue~1 & $23$ & $220.48,\: 50.43$ & $19.36^{+0.32}_{-0.35}$ & $13-37$ & $158$ & $0.06 - 100$ & $0.12$ & $-0.5$ \\
Ursa~Major~II & $32$ & $152.46,\: 37.44$ & $19.42^{+0.44}_{-0.42}$ & $35-45$ & $95$ & $0.12 - 100$ & $0.30$ & $-2.1$ \\
\hline
\hline
\end{tabular}
\caption{Summary of the dSph properties and observations by the MAGIC telescopes. We report the heliocentric distance and Galactic coordinates of each dSph, as well as the total \textit{J}-factor values and its $\pm 1\sigma$ uncertainties from GS15~\cite{2015ApJ...801...74G} used in the present work. We also report the zenith distance (Zd) range, the total observation time ($ T_{\text{obs}} $), and the energy range (E). We then list the angular radius ($\theta$) of the signal region, the normalization between background and signal regions ($ \tau $), and the significance of detection (S$_{\text{Li\&Ma}}$) calculated by following Li\&Ma~\cite{LiMa:1983}. Note that the significance of detection is not reported for Coma~Berenices and Draco in~\cite{2022PDU....3500912A}, but no gamma-ray excess have been found in any of these sources.}
\label{Tab:dSph}
\end{table}

\section{Likelihood analysis method}
\label{sec:LklAnalysis}

The low-level data of the four dSph observations (see Sec. \ref{sec:Observation}) were reduced by the MAGIC Collaboration using the standard MAGIC analysis software MARS~\citep{2013ICRC...33.2937Z} and published in~\citep{2022PDU....3500912A}. We re-analysed the high-level data products (event lists for \texttt{gLike}; likelihood curves for \texttt{LklCom}) in the context of brane-world extra-dimensional theories using the open-source analysis software tools~\citep{2021arXiv211201818M} for multi-instrument and multi-target DM searches \texttt{gLike}~\citep{Rico:gLike} and \texttt{LklCom}~\citep{Miener:LklCom}.

We followed the likelihood analysis scheme proposed by Aleksi\'{c}, Rico and Martinez in~\cite{2012JCAP...10..032A}, which is the standard DM analysis framework within the MAGIC Collaboration~\citep{2014JCAP...02..008A,2016JCAP...02..039M,2018JCAP...03..009A,2022PDU....3500912A,Rico:2020vlg}. Our final joint likelihood function $ \mathcal{L} $ is a nested product of the binned likelihood function for each dSphs ($N_{\text{dSphs}} = 4$) and their distinct observational datasets ($N_{\text{obs},k}$ refers to the number of datasets of dSph $ k $) with their corresponding set of instrument response functions (IRFs) caused by different observational conditions or hardware setup of the instrument. It reads for all datasets $ \bm{\mathcal{D}} $ with nuisance parameters $ \bm{\nu} $ as
\begin{equation}
    \label{eq:Lkl_function}
    \begin{split}
    \mathcal{L}\left( \langle \sigma v \rangle ; \bm{\nu} \mid \bm{\mathcal{D}} \right) = \prod_{k=1}^{N_{\text{dSphs}}} \Big\{ \prod_{l=1}^{N_{\text{obs},k}} \Big[ \prod_{i=1}^{N_{\text{bins}}} \Big( \mathcal{P} (s_{kli}&(\langle \sigma v \rangle) + b_{kli} \mid N_{\text{ON},kli}) \cdot \mathcal{P} (\tau_{kl} b_{kli} \mid N_{\text{OFF},kli}) \Big) \Big]  \\ 
    & \times \mathcal{T}_{kl} \left( \tau_{kl} \mid \tau_{\text{o},kl}, \sigma_{\tau_{kl}} \right) \Big\} \times \mathcal{J}_{k} \left( J_{k} \mid J_{\text{o},k}, \sigma_{\log_{10}J_{k}} \right)
    \end{split}
\end{equation}
\noindent
where $ \mathcal{P} (x | N) $ is the Poisson distribution of mean $x $ and measured value $ N $, $ s_{kli}(\langle \sigma v \rangle) $ and $ b_{kli} $ are the expected numbers of signal and background events in the $ i $-th energy bin, respectively, and $ N_{\text{ON},kli} $, $ N_{\text{OFF},kli} $ are the total number of observed events in a given energy bin $ i $ of the $ l $-th distinct dataset of the $ k $-th dSph in the signal (ON) and background (OFF) regions, respectively. Besides $ b_{kli} $, the normalization between background and signal regions $ \tau_{kl} $, described by the likelihood function $ \mathcal{T}_{kl} $\footnote{The likelihood function $ \mathcal{T}_{kl} \left( \tau_{kl} \mid \tau_{\text{o,kl}}, \sigma_{\tau_{kl}} \right) $ of the $ l $-th distinct dataset of the $ k $-th dSph is a Gaussian with mean $ \tau_{\text{o,kl}} $ and variance $ \sigma_{\tau_{kl}}^{2} $, which include statistical and systematics uncertainties. We consider a systematic uncertainty of $ \sigma_{\tau_{\mathrm{syst}}} = 1.5\%  \cdot \tau_{kl} $ on the estimate of the residual background based on a dedicated performance study of the MAGIC telescopes~\citep{2016APh....72...76A}.}, is a nuisance parameter in the analysis~\cite{2022PDU....3500912A}. We treat also the \textit{J}-factors as a nuisance parameters using the likelihood $ \mathcal{J}_{k} $ for the \textit{J}-factor of the $ k $-th dSph following~\citep{2015PhRvL.115w1301A}. In the absence of a branon DM signal, upper limits (ULs) on $\langle \sigma v \rangle$ are set using a test statistic following~\citep{2015PhRvL.115w1301A}.

\section{Results and Outlook}

\begin{figure}[ht]
    \centering
    \begin{subfigure}{.49\textwidth}
        \includegraphics[width=\textwidth]{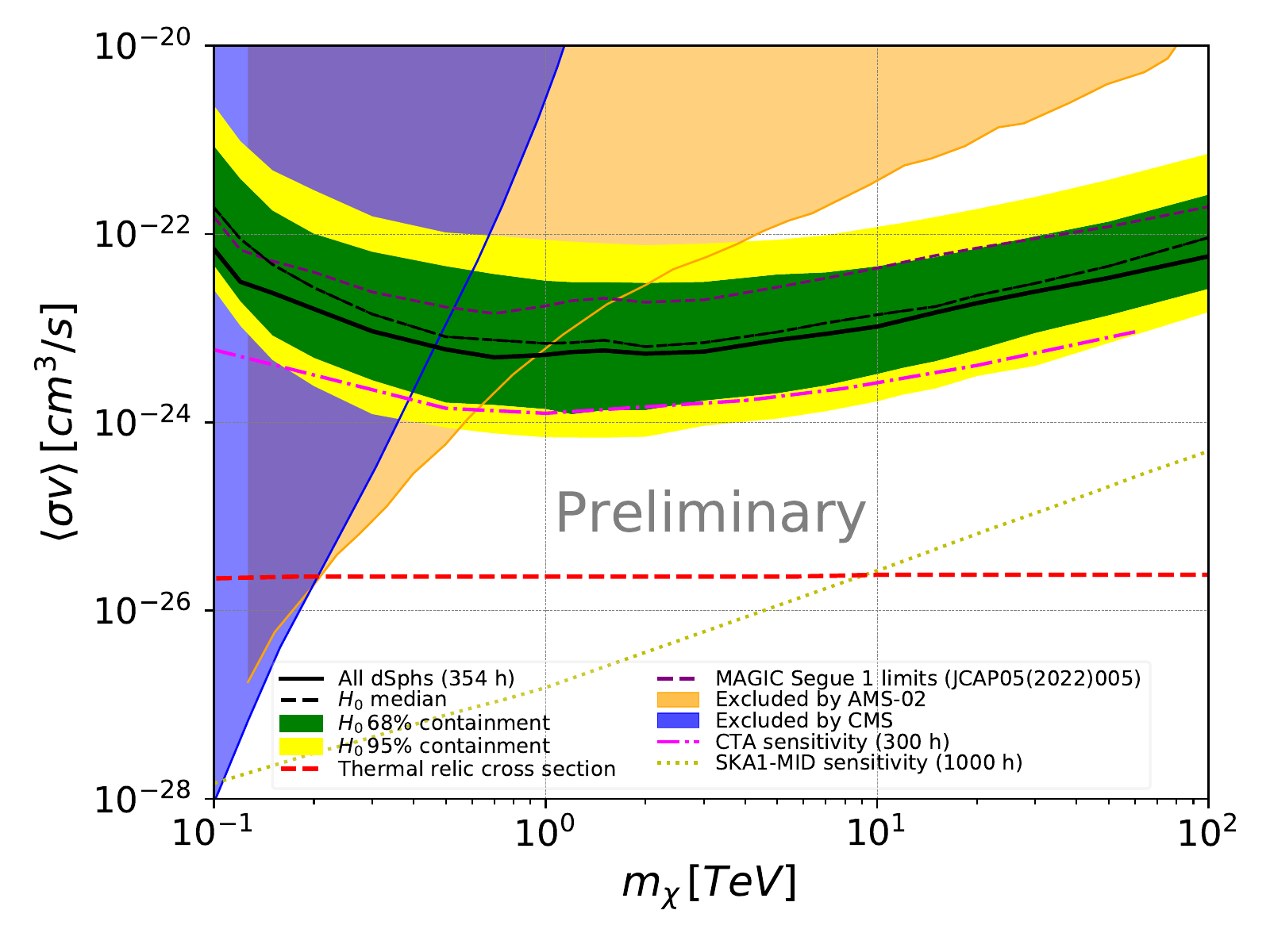}
    \end{subfigure}
    \begin{subfigure}{.49\textwidth}
        \includegraphics[width=\textwidth]{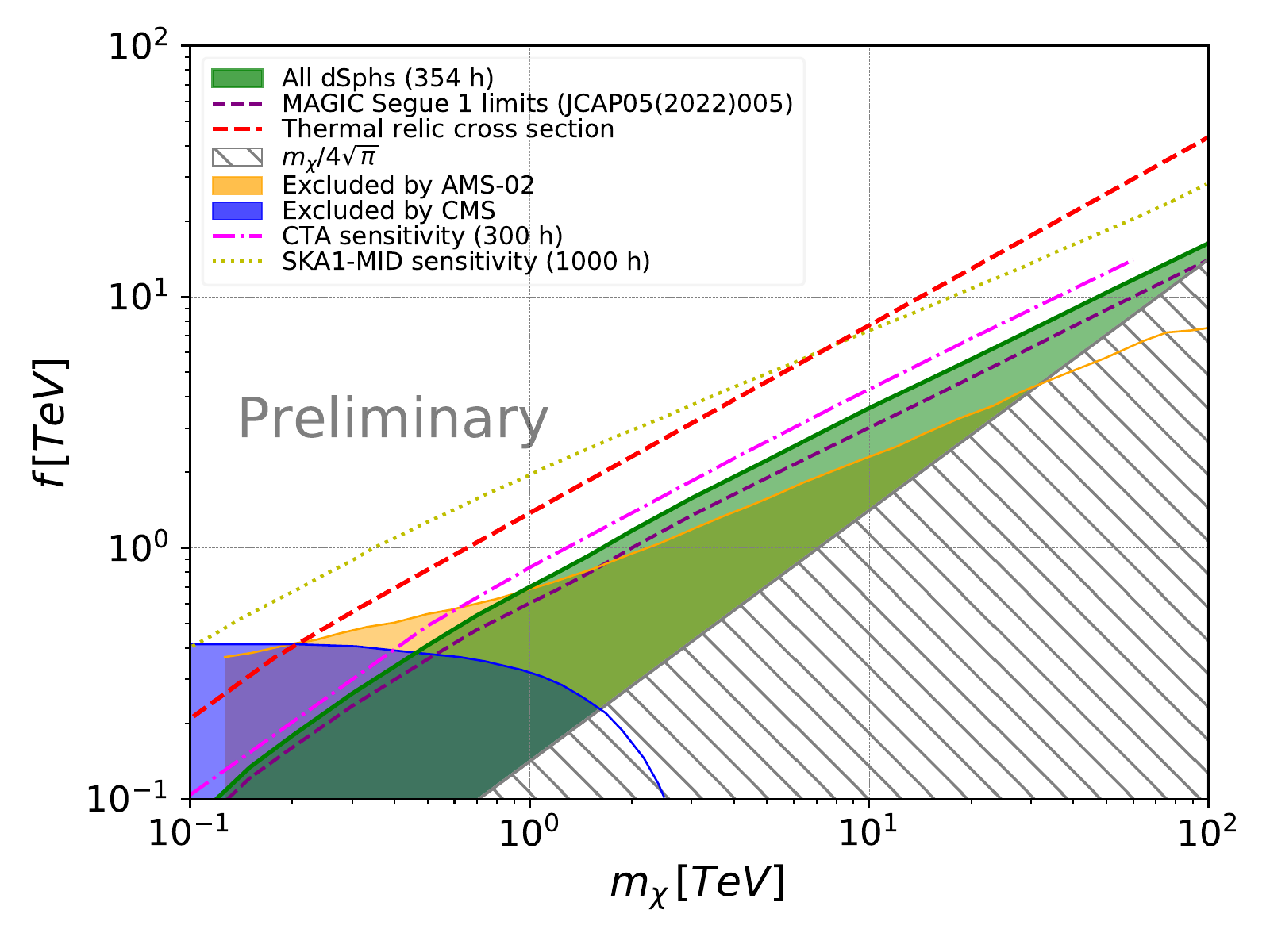}
    \end{subfigure}
\caption{$ 95 \% $ confidence level ULs on $ \langle \sigma v \rangle $ (left panel) and on $ f $ (right panel) for branon DM annihilation from the combined analysis of 354 hours of dSph observations. See text for more details.}
\label{fig:Limits}
\end{figure}

We present the observational $ 95 \% $ confidence level ULs on the thermally-averaged cross-section $ \langle \sigma v \rangle $ and on the brane tension $ f $ (see further details in~\cite{2022JCAP...05..05}) for branon DM annihilation obtained with 354 hours of dSph observations by the MAGIC telescopes (see Fig. \ref{fig:Limits})). We perform a multi-target search for branon DM particles of masses between $ \unit[100]{GeV} $ and $ \unit[100]{TeV} $. As expected from the no significant gamma-ray excess in the dSph observations by the MAGIC telescopes~\cite{2022PDU....3500912A}, our constraints for branon DM annihilation are located within the $ 68 \% $ containment band, which is consistent with the no-detection scenario.

This work leads to the most constraining branon DM limits in the TeV mass range, superseding previous constraints by CMS~\cite{2016physletb.755.102} (blue), AMS-02~\cite{2019physletb.790.345} (orange) and MAGIC limits for Segue1 alone~\cite{2022JCAP...05..05,2022arXiv220507055M} (purple). The prospects of the future CTA~\cite{2020JCAP...10..041A} (magenta) and SKA~\cite{2020PDU....2700448C} (yellow) are also depicted in Fig.~\ref{fig:Limits}. We obtain our strongest limit $ \langle \sigma v \rangle \simeq \unit[4.9 \times 10^{-24}]{cm^{3}s^{-1}} $ for a $ \sim \unit[0.7]{TeV} $ branon DM particle mass. We can achieve even more stringent and robust exclusion limits by adding further dSph observations of the MAGIC telescopes or other gamma-ray~\cite{Oakes:2019,Armand:2021} or neutrino telescopes to this analysis scheme.

\tiny

\section*{Acknowledgments}
\scriptsize We would like to thank the Instituto de Astrofísica de Canarias for the excellent working conditions at the Observatorio del Roque de los Muchachos in La Palma. The financial support of the German BMBF, MPG and HGF; the Italian INFN and INAF; the Swiss National Fund SNF; the ERDF under the Spanish Ministerio de Ciencia e Innovación (MICINN) (FPA2017-87859-P, FPA2017- 85668-P, FPA2017-82729-C6-5-R, FPA2017-90566-REDC, PID2019-104114RB-C31, PID2019-104114RB-C32, PID2019- 105510GB-C31C42, PID2019-~107847RB-C44, PID2019-107988GB-C22); the Indian Department of Atomic Energy; the Japanese ICRR, the University of Tokyo, JSPS, and MEXT; the Bulgarian Ministry of Education and Science, National RI Roadmap Project DO1-268/16.12.2019 and the Academy of Finland grant nr. 317637 and 320045 are gratefully acknowledged. This work was also supported by the Spanish Centro de Excelencia “Severo Ochoa” SEV-2016- 0588, SEV-2017-0709 and CEX2019-000920-S, and "María de Maeztu” CEX2019-000918-M, the Unidad de Excelencia “María de Maeztu” MDM-2015-0509-18-2 and the "la Caixa" Foundation (fellowship LCF/BQ/PI18/11630012), by the Croatian Science Foundation (HrZZ) Project IP-2016-06-9782 and the University of Rĳeka Project 13.12.1.3.02, by the DFG Collaborative Research Centers SFB823/C4 and SFB876/C3, the Polish National Research Centre grant UMO-2016/22/M/ST9/00382 and by the Brazilian MCTIC, CNPq and FAPERJ.\newline
TM acknowledges support from PID2019-104114RB-C32.\newline
VG’s contribution to this work has been supported by Juan de la Cierva-Incorporaci\'on IJC2019-040315-I grant, and by the PGC2018-095161-B-I00 and CEX2020-001007-S projects, both funded by MCIN/AEI/10.13039/501100011033 and by "ERDF A way of making Europe". VG thanks J.A.R. Cembranos for useful discussions. \newline
DK is supported by the European Union's Horizon 2020 research and innovation programme under the Marie Sk\l{}odowska-Curie grant agreement No. 754510. DK and JR acknowledge the support from MCIN/AEI/ 10.13039/501100011033 under grants PID2019-107847RB-C41 and SEV-2016-0588 ("Centro de Excelencia Severo Ochoa"), and from the CERCA institution of the Generalitat de Catalunya.
\\
\\
This paper has gone through internal review by the MAGIC Collaboration.

\end{document}